\documentstyle[aps,prc,twocolumn,floats,epsfig]{revtex}
\draft
\renewcommand{\vec}[1]{{\mathbf{#1}}}
\begin{document}

\renewcommand{\thefootnote}{\arabic{footnote}}

\twocolumn[\columnwidth\textwidth\csname@twocolumnfalse\endcsname

\title{Potential Energy Surfaces of Superheavy Nuclei}

\author{M. Bender${}^{\rm 1,2,}$\cite{NewAddress},
        K. Rutz${}^{\rm 1,3}$,
        P.--G. Reinhard${}^{\rm 2,4}$, 
        J. A. Maruhn${}^{\rm 1,4}$,
        and W. Greiner${}^{\rm 1,4}$}

\address{${}^{\rm 1}$Institut f\"ur Theoretische Physik, 
         Universit\"at Frankfurt,
         Robert--Mayer--Str.~10, D--60325 Frankfurt am Main, Germany.}

\address{${}^{\rm 2}$Institut f\"ur Theoretische Physik II, 
         Universit\"at Erlangen--N\"urnberg,
         Staudtstr.~7, D--91058 Erlangen, Germany.}

\address{${}^{\rm 3}$Gesellschaft f\"ur Schwerionenforschung mbH,
         Planckstr.~1, D--64291 Darmstadt, Germany.}

\address{${}^{\rm 4}$Joint Institute for Heavy--Ion Research, 
         Oak Ridge National Laboratory,
         P.~O.\ Box 2008, Oak Ridge, TN~37831.}

\date{\today}

\maketitle

\addvspace{5mm}

%
%========================================================================
%
\begin{abstract}
We investigate the structure of the potential energy surfaces of the
superheavy nuclei ${}^{258}_{158}{\rm Fm}_{100}$,
${}^{264}_{156}{\rm Hs}_{108}$, ${}^{278}_{166}{112}$,
${}^{298}_{184}{114}$, and ${}^{292}_{172}{120}$
within the framework of selfconsistent nuclear models, i.e.\ the
Skyrme--Hartree--Fock approach and the relativistic mean--field model.
We compare results obtained with one representative parametrisation 
of each model which is successful in describing superheavy nuclei.
We find systematic changes as compared to the potential energy
surfaces of heavy nuclei in the uranium region: there is no 
sufficiently stable
fission isomer any more, the importance of triaxial configurations to
lower the first barrier fades away, and asymmetric fission paths
compete down to rather small deformation. Comparing the two models, it
turns out that the relativistic mean--field model gives generally
smaller fission barriers.
\end{abstract}
\pacs{PACS numbers: 
      21.30.Fe % Forces in hadronic systems and effective interactions
      21.60.Jz % Hartree-Fock and random-phase approximations
      24.10.Jv % Relativistic models
      27.90.+b % 220 <_ A
     }

\addvspace{5mm}]

\narrowtext
%
%========================================================================
%
\section{Introduction}
Superheavy nuclei are by definition those nuclei with charge numbers beyond
the heaviest long--living nuclei that have a negligible liquid--drop 
fission barrier, i.e.\ they are only stabilized by shell effects 
\cite{SuperNils,Mosel}.
The stabilizing effect of the shell structure has been demonstrated
in recent experiments at GSI \cite{Z111,Z112} and Dubna \cite{Dubna}, 
where an island of increased stability in the
vicinity of the predicted doubly magic deformed nucleus
${}^{270}_{162}{\rm Hs}$ \cite{Pat91a,Mol92a,Mol94a} has been reached.

The full potential energy surface (PES) of superheavy nuclei is of
interest as it allows to estimate the stability against spontaneous
fission and to predict the optimal fusion path for the synthesis 
of these nuclei. Both features are of great importance for planning 
future experiments. There are numerous papers on the structure of the
potential--energy surfaces of superheavy nuclei in macroscopic--microscopic
models, see, e.g., \cite{Mol94a,Cwi92a,Smo95a},
but only very few investigations in selfconsistent models so far.
A systematic study of the deformation energy of superheavy nuclei
along the valley of $\beta$ stability in the region 
\mbox{$100 \leq Z \leq 128$} and \mbox{$150 \leq N \leq 218$}
in HFB calculations with the Gogny force $D1s$ under restriction to axially
and reflection symmetric shapes was presented in \cite{Ber96a}.
The full potential energy surface in the $\beta$--$\gamma$ plane 
of a few selected nuclei as resulting from Skyrme--Hartree--Fock 
calculations in a triaxial representation is discussed in \cite{Cwi96a}. 
This investigation stresses the importance of non--axial shapes, that 
lower the fission barrier of some superheavy nuclei to half its value 
assuming axial symmetry.
There is still no selfconsistent calculation of the deformation energy
of superheavy nuclei allowing for reflection--asymmetric shapes.

In a series of papers we have demonstrated the uncertainties in the 
extrapolation of the shell structure to the region of superheavy nuclei 
within selfconsistent models \cite{SHsphere,SHdef}. The reasons
for the different behavior of parametrisations that work comparably
well for conventional stable nuclei when extrapolated to large
mass numbers can be traced to differences in the effective mass and the 
isospin--dependence of the spin--orbit interaction \cite{SHshells}.
It is the aim of this paper to investigate the important degrees of 
freedom of the potential energy surface of superheavy nuclei for the
example of a few selected nuclides, i.e.\ ${}^{258}_{158}{\rm Fm}_{100}$,
${}^{264}_{156}{\rm Hs}_{108}$, ${}^{278}_{166}{112}$,
${}^{298}_{184}{114}$, and ${}^{292}_{172}{120}$
within the framework of selfconsistent nuclear structure models, namely
the relativistic mean--field model (RMF, for reviews see 
\cite{WalSer,Rei89}) and the nonrelativistic Skyrme--Hartree--Fock (SHF)  
approach (for an early review see \cite{refSHF}),
in both cases including also reflection--asymmetric shapes.
%
%========================================================================
%
\section{The framework}
The comparison of the calculated binding energies 
of the heaviest known even--even nuclei with the experimental values
\cite{SHsphere,SHdef} has shown that 
the Skyrme parametrisation SkI4 and the relativistic force \mbox{PL--40}
are to be among the preferred parametrisations for the extrapolation to
superheavy nuclei.
The nonrelativistic force SkI4 is a variant of the Skyrme 
parametrisation where the spin--orbit force is complemented 
by an explicit isovector degree--of--freedom \cite{SkIx}. 
The energy functional and the parameters are presented in 
Appendix~\ref{Subsect:SHF}.
The modified spin--orbit force has a strong effect on the spectral 
distribution in heavy nuclei and produces a big improvement 
concerning the binding energy of superheavy nuclei \cite{SHsphere,SHdef}. 
The RMF parametrisation \mbox{PL--40} \cite{PL40} aims at a best 
fit to nuclear 
ground--state properties with a stabilized form of the scalar nonlinear 
selfcoupling, see Appendix~\ref{Subsect:RMF} for details.
It shares most properties with the widely used standard 
nonlinear force NL--Z \cite{NLZ}.

Both models are implemented in a common framework sharing
all the model--independent routines. The numerical procedure represents the 
coupled SHF and RMF equations on a grid in coordinate space using a 
Fourier definition of the derivatives and solves them with the damped 
gradient iteration method \cite{dampgrad}. An axial representation 
allowing 
for reflection--asymmetric shapes is employed in most of the calculations, 
while a triaxial deformed representation is used to investigate the 
influence of non--axial configurations on the first barrier.

In both SHF and RMF the pairing correlations are treated in the BCS
scheme using a delta pairing force \cite{Krieger} \mbox{$V_{\rm pair}
= V_{\rm p/n} \, \delta(\textbf{r}_1 - \textbf{r}_2)$},
see Appendix~\ref{Subsect:PairFunc} for details.
This pairing force has the technical advantage that the
strengths $V_{\rm p/n}$ are universal numbers which hold throughout
the chart of nuclei, different from the widely used seniority model,
where the strengths need to be parametrised with $A$ dependence, and
therefore in the description of a fission process would have to be
interpolated between the values for the initial nucleus and averaged
values for the fission fragments. 

Furthermore, a center--of--mass correction is performed by subtracting 
a posteriori 
\mbox{$E_{\rm c.m.}=\langle\hat\textbf{P}\rule{0pt}{6.8pt}^2_{\rm c.m.}
\rangle/(2mA)$},
see \cite{Rei89,SkyrmeFit}, as done in the original fit of the 
parametrisations. This treatment of the center--of--mass correction 
is a fair approximation, its uncertainty for the heavy systems discussed  
here is smaller than \mbox{$0.2 \; {\rm MeV}$} \cite{cmmotion}.
The center--of--mass correction, however, has to be complemented by
corrections for spurious rotational and vibrational modes as well. Their
proper implementation is a very demanding task, as it requires the 
appropriate cranking masses. As done in most other mean--field 
calculations,
we omit this detail. In the barrier heights, which we will discuss here,  
only the variation of these corrections with deformation enters. 
An estimate for these effects can be taken from a two--center shell model
calculation of actinide nuclei \cite{ZPE,GCMRev}:
the amplitude of the corrections increases with
increasing deformation, lowering the first barrier by approximately
\mbox{$0.5 \; {\rm MeV}$} and the second barrier by \mbox{$2 \; {\rm MeV}$}.
There is an uncertainty due to the numerical solution
of the equations of motion which is of the order of \mbox{$0.1 \; {\rm MeV}$}
even for large deformations, thus negligible in our calculations.
The prescription of pairing adds another uncertainty to the calculated
binding energies. We use the same pairing scheme and force for all
calculations with an optimized strength for each mean--field
parametrisation. The use of a local pairing force improves
the description of pairing correlations within the BCS scheme 
compared to a constant force or constant gap approach \cite{pairStrength},
and removes some problems concerning the coupling of continuum states to  
the nucleus. From possible variation of pairing recipes, we
assume an uncertainty of the total binding energy of approximately 
$1 \; {\rm MeV}$ \cite{ReiLN}.

In the following, we will present deformation energy curves calculated 
with a quadrupole constraint (for numerical details see 
\cite{Asymrmf}). In a constrained selfconsistent calculation all 
unconstrained multipole deformations (of protons and neutrons separately)
are left free to adjust themselves to a minimum energy configuration
within the chosen symmetry. 
Thus the selfconsistent description of the potential--energy surface
takes many more degrees of freedom into account than the $3$-$5$ shape
parameters that can be handled within macroscopic--microscopic
calculations. The macroscopic--microscopic models have the additional
technical disadvantage that, for the description of a fission process,
several nucleon--number dependent terms in both the parametrisation of
the macroscopic and the microscopic model have to be interpolated
between the values for the compound system and the fragments (see,
e.g., \cite{Mol94a}), leading to an uncertainty of the binding energy
in the intermediate region.
It is to be noted, however, that there remains some open end
concerning shapes also in the selfconsistent models as there might
exist several local minima which are separated by a potential
barrier. The numerical procedure solving the constrained mean--field
equations converges usually to the next local minimum, depending on
the initial state. And it requires experience as well as patient
searches to make sure that one has explored all local minima in a
given region.

The deformation energy curves presented in the following are shown versus 
the dimensionless multipole moments of the mass density which are
defined as
\begin{displaymath}
\beta_\ell
= \frac{4 \pi}{3 A r_0^\ell} \; \langle r^\ell \, Y_{\ell 0} \rangle 
\quad \hbox{with $r_0 = 1.2 \, A^{1/3} \; $fm.}
\end{displaymath}
Note that these $\beta_\ell$ are computed as expectation values from the  
actual mass distribution of the nucleus and need to be distinguished
from the generating deformation parameters which are used in the 
multipole expansion of the nuclear shape in macroscopic models 
\cite{beta}. Besides the description in terms of $\beta_\ell$, we 
will indicate the various shapes along the paths in all figures, by 
the mass density contours at \mbox{$\rho_0 = 0.07 \, {\rm fm}^{-3}$}. 
Furthermore, when looking at potential energy surfaces, one
should keep in mind that these are only the first indicators of the fission
properties. A more detailed dynamical description requires also the 
knowledge
of the collective masses along the path. This is, however, a very
ambitious task which goes beyond the aim of this contribution.
We intend here mainly a qualitative discussion of the potential
landscape.
%
%========================================================================
% 
\section{Results and Discussion} 
\begin{figure}[t!]
\epsfig{file=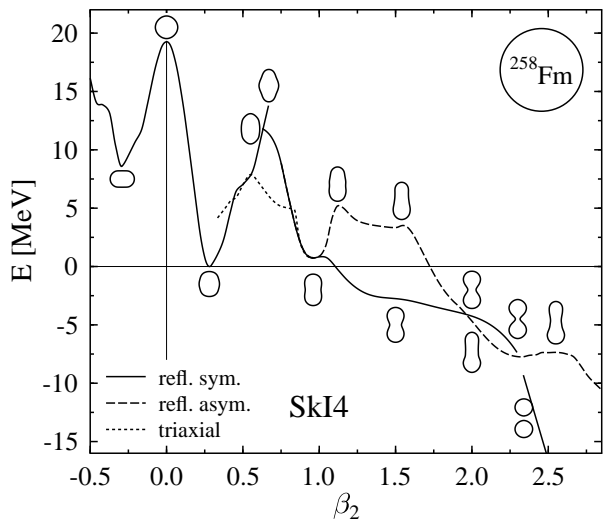}
\caption{\label{fm258ski4}
Valleys in the PES of ${}^{258}{\rm Fm}$ for SkI4 from calculations
in axial symmetry with (``refl. sym.'') and without (``refl. asym.'')
reflection symmetry. In the vicinity of the first barrier also the
result from a non--axial calculation (``triaxial'') is shown. To give
an impression of the nuclear shapes along the path, mass density
contours at \mbox{$\rho_0=0.07 \; {\rm fm}^{-3}$} are drawn near the
corresponding curves.
}%
\end{figure}

Figure~\ref{fm258ski4} shows results of a Skyrme--Hartree--Fock 
calculation 
with SkI4 for ${}^{258}{\rm Fm}$, a nucleus that is located at the lower
end of the region of superheavy nuclei. 
%
%===========
%
%
%==========
%%
The strong shell effect in the prolate ground state lowers 
the binding energy of this nucleus by \mbox{$19.3 \; {\rm MeV}$} or $1 \%$ 
compared to a spherical shape, demonstrating the importance of
considering deformations for the calculation of the ground--state
binding energies in this region of the chart of nuclei.
The first barrier is lowered from 11.8 MeV to 7.7 MeV when allowing for
triaxial configurations. But the preference for triaxial shapes at the
top of the barrier disappears when going to both larger and smaller
deformations. It is interesting to note that the axial solutions are not
continuously connected from ground state through first minimum
when using a constraint on the quadrupole moment, but
develop in two branches distinguished by their hexadecapole moment.
The ground--state branch has a diamond like shape with a $\beta_4$ much
larger than the branch coming from outside. The continuous connection
is established by the intermediate triaxial shapes.
%
%=======
%
\begin{figure}[t!]
\epsfig{file=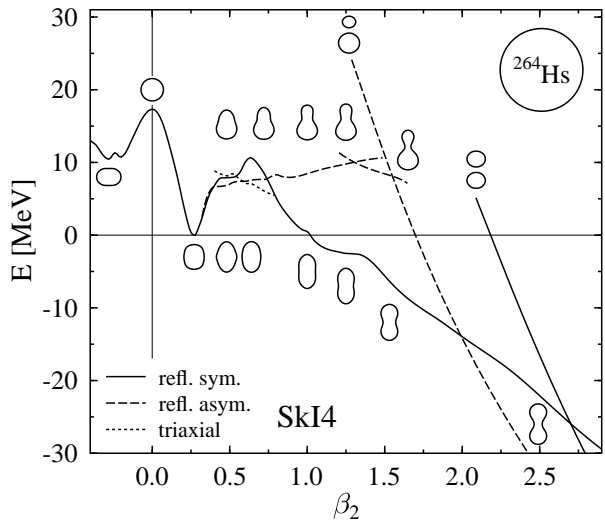}
\caption{\label{hs264ski4}
  Valleys in the PES of ${}^{264}{\rm Hs}$ for SkI4, drawn in the same
  manner as in Fig.~\protect{\ref{fm258ski4}}.
}
\end{figure}
%
%=======
%
%
%=======
%
\begin{figure}[b!]
\epsfig{file=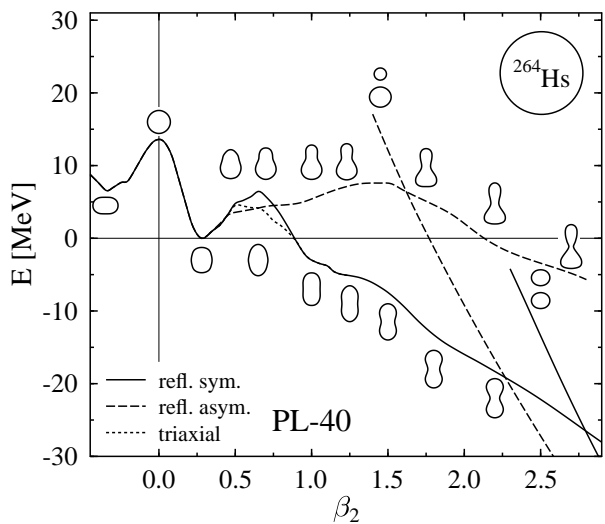}
\caption{\label{hs264pl40}
  Valleys in the PES of ${}^{264}{\rm Hs}$ for \mbox{PL--40}, 
  drawn in the same manner as in Fig.~\protect{\ref{fm258ski4}}.
}
\end{figure}
%
%=======
%

The PES of ${}^{258}{\rm Fm}$ shows some significant deviations from the  
familiar double--humped fission barrier of the somewhat lighter
nuclei in the plutonium region \cite{dblhump}.
There is no superdeformed minimum in the PES that can be associated with a
fission isomer because the second barrier vanishes in case of symmetric 
breakup. This is due to the strong shell effect of the closed spherical 
\mbox{$Z = 50$} shell in the two fragments, that reaches far inside to 
deformations as small as \mbox{$\beta_2=1.0$}, the usual location of
the fission  
isomer. This is reflected in the evolution of shapes along the symmetric
path, that look like two intersecting spheres.
At large deformations around \mbox{$\beta_2\approx 1.5$} a valley with
finite mass asymmetry appears, which is separated from the symmetric
valley by a 
small potential barrier, but \mbox{$5 \; \mbox{MeV}$} higher in energy. The
occurrence of competing but well  separated valleys and the consequences
for fission or fusion complies with the results from macroscopic models,
for a discussion see e.g.\ \cite{Mol94a,Cwi89a}.

Figure~\ref{hs264ski4} shows the valleys in the PES of ${}^{264}{\rm Hs}$,
at present the heaviest known even--even nucleus \cite{Z108}.
Although the fragments from a symmetric breakup of ${}^{264}{\rm Hs}$ are
far from any shell closure, this channel of the PES keeps the 
characteristic
structure of the PES of ${}^{258}{\rm Fm}$ like the absence of a fission
isomer and the vanishing second barrier. The fission path will follow the 
reflection symmetric solution, that gives a much narrower barrier than 
the asymmetric solution. Although the first barrier has similar width 
and height as the first barrier of typical actinide nuclei, the absence 
of the second barrier will lower the lifetime against spontaneous fission
dramatically. Like in the actinide region, the first barrier is a bit
lowered if one allows for triaxial configurations. The reflection
asymmetric solution does not lower the overall barrier, but it
coexists far inside the barrier. This is a new feature occurring in the PES
of many superheavy nuclei and was already found in macroscopic--microscopic
calculations in the two--center shell model \cite{Sandu}.
The asymmetric path connects the asymptotically separated combination 
\mbox{${}^{210}{\rm Po} + {}^{54}{\rm Cr}$} with the ground 
state, and corresponds to the fusion path. This combination of
projectile and target differs only slightly from the experimentally
successful choice \mbox{${}^{207}{\rm Pb}({}^{58}{\rm Fe},{\rm n}){}^{264}{\rm
Hs}$} \cite{Z108}. It reflects the strong shell effect of an asymmetric
breakup with a heavy fragment in the region of doubly magic
${}^{208}{\rm Pb}$. It is interesting to compare that with the case of
actinide nuclei: these have also a well developed asymmetric path which
is, however, confined to large deformations and reaches only down to the
outer barrier \cite{Asymrmf}.

A calculation of the PES of ${}^{264}{\rm Hs}$ in the RMF using 
\mbox{PL--40} 
gives qualitatively the same results, see Fig.\ \ref{hs264pl40}, but 
there are some differences in details. The barrier is a bit smaller
in height and width than for SkI4. The lowering of the barrier 
for \mbox{PL--40} is due to the smaller shell effect for
the ground--state configuration in this parametrisation.
While for SkI4 this nucleus has a deformed proton magic number,
\mbox{PL--40} does not predict a shell closure for \mbox{$Z = 108$} at all,
see \cite{SHdef}. The effect of non--axial configurations on the height
of the barrier is of the same size as in SkI4.
%
%=======
%
\begin{figure}[t!]
\epsfig{file=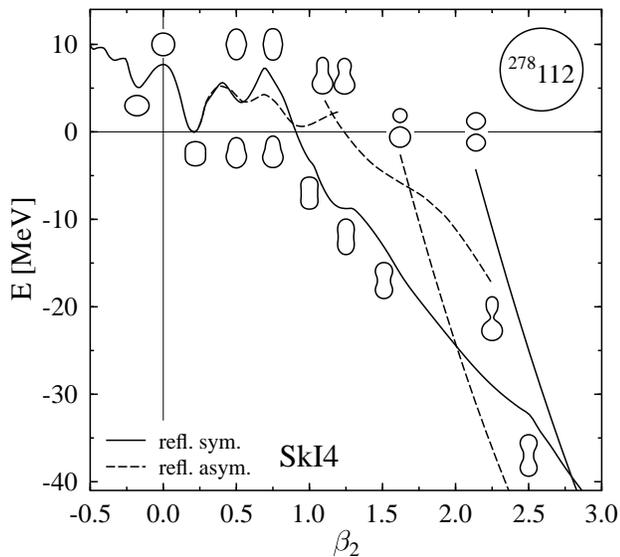}
\caption{\label{166112ski4}
  Valleys in the PES of ${}^{278}_{166}{\rm 112}$ for SkI4, 
  drawn in the same manner as in Fig.~\protect{\ref{fm258ski4}}.
}
\end{figure}
%
%=======
%

As an example for a nucleus located at the upper border of the known chart
of nuclei, Fig.\ \ref{166112ski4} shows the valleys in the PES
of ${}^{278}_{166}{\rm 112}$, calculated with SkI4.
This nuclide corresponds to the compound nucleus in the cold fusion
reaction ${}^{208}{\rm Pb}({}^{70}{\rm Zn},{\rm n}){}^{277}{112}$
which was used to synthesize the heaviest detected superheavy nucleus
so far \cite{Z112}.
Although the proton number of this nucleus is close to the value 
\mbox{$Z = 114$} for the next spherical proton shell closure predicted by
SkI4, its neutron number is quite far from the next predicted spherical 
neutron shell closure \mbox{$N = 184$} but close to the deformed shell  
closure 
\mbox{$N = 162$}, that drives the nucleus to a strong prolate deformation 
with \mbox{$\beta_2 = 0.22$}, \mbox{$\beta_4 = - 0.09$}, see \cite{SHdef} 
for details.
The PES of ${}^{278}_{166}{\rm 112}$ shares most
overall features with the PES of ${}^{264}{\rm Hs}$, like the 
one--humped structure and the lowering of the first barrier due to 
asymmetric configurations, but there are some differences in detail. The
(symmetric) fission barrier is narrower and slightly smaller (7.3 MeV
compared with 10.6 MeV) than for ${}^{264}{\rm Hs}$. We have checked the
effect of triaxial shapes and found that they are not effective to lower
the first barrier for this superheavy nucleus. 
At superdeformed shapes \mbox{$\beta_2 \approx 0.5$} a spurious minimum 
develops in the symmetric barrier, that in reality is a saddle point, 
since the potential drops for asymmetric deformations.
Note that the barrier is very soft in mass asymmetry in this region.
Even at quadrupole deformations as small as 
\mbox{$\beta_2 = 0.5$} the binding energy is nearly
constant within the range \mbox{$0 < \beta_3 < 0.3$}. The asymmetric
path shows a rich substructure. There is a shallow minimum at 
\mbox{$\beta_2 \approx 0.9$}, while around \mbox{$\beta_2 = 1.2$}, 
the results show a
transition between two solutions with slightly different hexadecapole
moment corresponding to shapes with differently pronounced ``necks'' but
nearly constant mass asymmetry. The asymmetric valley corresponds to the
breakup \mbox{${}^{210}{\rm Po}+{}^{68}{\rm Ni}$}, that is quite close to the
projectile--target combination used for the synthesis of this nuclide.

Axial and reflection symmetric calculations within the semi--microscopic
``extended Thomas--Fermi--Strutinski integral method'' (ETFSI) \cite{ETFSI}
that uses a Skyrme force, i.e.\ SkSC4, for the nuclear interaction as 
well, found superdeformed minima in the PES of this and many other
superheavy nuclei in the region \mbox{$Z \geq 112$} with 
\mbox{$\beta_2 \approx 0.45$} and a larger binding energy than the
usual minima at small deformations. 
Our results indicate that these superdeformed minima vanish or 
will have a rather small fission barrier when reflection--asymmetric
shapes are taken into account. Therefore the usual minimum in the PES at
smaller $\beta_2$ has to be considered as the ground--state configuration,
having a still sizeable first barrier and thus the larger fission
half--life as compared to the competing minimum.

All nuclei discussed so far are located in the region of known
superheavy nuclei. Now we want to look at possible candidates for the
spherical doubly magic superheavy nucleus. As shown in 
\cite{SHsphere,SHdef,SHshells}, the predictions for doubly magic nuclei 
differ significantly between SHF and RMF. The RMF predicts 
${}^{292}_{172}120$ to be doubly magic, while
the extended Skyrme force SkI4 prefers ${}^{298}_{184}114$, the
nucleus that has been predicted to be the center of the island of
superheavy nuclei for a long time \cite{SuperNils,Mosel}. 
Other Skyrme forces, however, do not predict any doubly magic spherical 
nuclei in this region at all or shift the center of the island of 
superheavy nuclei to ${}^{310}_{184}126$, for example the force SkP 
\cite{SHsphere}. The PES of this nucleus, calculated with SkP allowing 
for triaxial shapes, is discussed 
in \cite{Cwi96a}. We now look at the two other candidates.
%
%=======
%
\begin{figure}[t!]
\epsfig{file=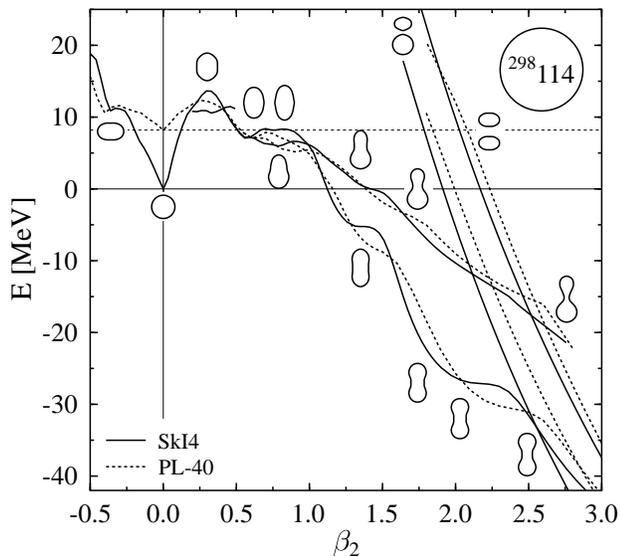}
\caption{\label{184114}
Valleys in the PES of ${}^{298}_{184}{114}$ for SkI4
and \mbox{PL--40}. Results from the calculations in different symmetries
can be distinguished by the mass density
contours which are drawn near the corresponding curves.
In the vicinity of the first barrier for SkI4 also the
result from a non--axial calculation is shown, that lowers
the barrier.
}
\end{figure}
%
%=======
%

Figure~\ref{184114} shows the paths of minimum potential energy in the PES 
of ${}^{298}_{184}114$, calculated with SkI4 (solid line) and \mbox{PL--40}
(dotted line). While \mbox{PL--40} shows only a weak neutron shell 
closure for this
nucleus, ${}^{298}114$ is the spherical doubly magic superheavy nucleus
predicted from SkI4. Therefore both forces lead to a spherical
ground state of this nucleus, but with differently pronounced shell 
effects. In the figure, the PES from calculations with \mbox{PL--40} 
is shifted with respect to SkI4 in such a way,
that the (spurious) shallow symmetric minimum at 
\mbox{$\beta_2 \approx 0.6$}
has the same energy in both models. For deformations larger than 
\mbox{$\beta_2 \approx 0.5$}, both forces coincide in their prediction 
for the PES: The second barrier vanishes if asymmetric shapes are 
taken into account,
but at large deformations the symmetric path is energetically favored.
Even the shell fluctuations that lead to steps in the symmetric path
are located at the same deformation for both forces. The significant 
difference between the potential energy surfaces occurs at small 
deformations \mbox{$\beta_2 < 0.5$}. The binding energy of the spherical
configuration, measured from the reference point is lowered by 
$ 7 \;{\rm MeV}$ for SkI4, but raised by approximately 
\mbox{$1.3 \; {\rm MeV}$}
for \mbox{PL--40}. Nevertheless the spherical configuration is the 
ground--state for both forces. It remains to be noted that around
\mbox{$\beta_2 \approx 0.3$} the barrier is slightly lowered for triaxial
shapes, for SkI4 by approximately \mbox{$3 \; {\rm MeV}$}, while for 
\mbox{PL--40} the gain in energy is only a few hundred keV.
%
%=======
%
\begin{figure}[t!]
\epsfig{file=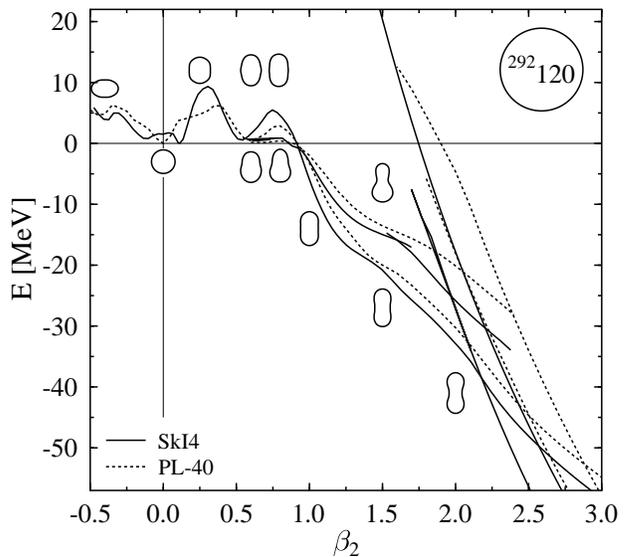}
\caption{\label{172120}
Valleys in the PES of ${}^{292}_{172}{120}$ for SkI4 and \mbox{PL--40}, 
drawn in the same manner as in Fig.~\protect{\ref{184114}}.
}
\end{figure}
%
%=======
%

As a final example, we consider the nucleus ${}^{292}_{172}{\rm 120}$
which has a spherical ground state and is a doubly magic system when
computed with \mbox{PL--40}. Figure~\ref{172120} shows its PES for 
\mbox{PL--40} and SkI4. The PES confirms the spherical minimum for 
\mbox{PL--40} whereas SkI4 prefers a
slightly prolate ground state. It is to be remarked, however,
that the actual ground state includes some quadrupole fluctuations 
around the minimum. In view of the weak deformation and small barrier
at zero deformation it requires a more elaborate calculation
including correlations to decide whether the true ground state
will be spherical or deformed.
A rather unexpected result is that the first barrier for \mbox{PL--40} is 
indeed much lower than that for SkI4.
This is a general result, that is also found for actinide nuclei
like ${}^{240}{\rm Pu}$. The fission half--lives from \mbox{PL--40} will  
thus be generally smaller and therefore ${}^{292}_{172}{\rm 120}$ will be 
more stable within SkI4 than with \mbox{PL--40} although the
latter predicts this as a doubly magic nucleus. Both forces
predict a strongly competing second minimum which, however, cannot
stabilize as a ground--state configuration (or serious isomer) because
the low second barrier makes it extremely unstable against fission.
This was already found in the previous examples and seems to be a
general feature of superheavy nuclei. The shell effects cease to
be strong enough to counterweight any more the strong decrease from
Coulomb repulsion.
%
%========================================================================
%
\section{Conclusions}
We have presented results of constrained selfconsistent calculations
of superheavy nuclei within the Skyrme--Hartree--Fock and 
relativistic mean--field model. 
The global structure of the PES of superheavy nuclei shows some 
significant differences compared to the well--known double--humped
fission barrier of heavy nuclei \mbox{$90 \leq Z < 100$}. The barrier of
superheavy nuclei is only single--humped. For the lighter superheavy
nuclei we still find that triaxial configurations lower the first
barrier, i.e., the region between the ground state and 
\mbox{$\beta_2 \approx 1.0$}. 
This effect vanishes for the heavier nuclei in the region
\mbox{$114 \leq Z \leq 120$} discussed here, but reappears in the 
heavier nuclei around ${}^{310}_{184}126$ \cite{Cwi96a}. The second
minimum (the fission isomer in the actinides) looses significance for
all superheavy nuclei because it becomes unstable against (asymmetric)
fission.
Seen from the reverse side, it turns out that the shell
structure of the final fragments influences the PES down to small
deformations. The  asymmetric channel with $^{208}$Pb as one fragment
thus carries through deep into the first barrier. This corresponds most
probably to the optimal fusion path whereas fission proceeds preferably
along the symmetric shapes. The global patterns of the paths are less
model dependent than for the actinides. Differences are most pronounced
in the vicinity of the ground states. They are caused by differences in
the detailed shell structure and lead to dramatically different
predictions for shell closures and fission half lives. The existence
and stability of superheavy nuclei is thus a most sensitive probe for
the present mean--field models.
%
%========================================================================
%
\acknowledgments
The authors would like to thank S.~Hofmann and G.~M\"unzenberg
for many valuable discussions.
This work was supported by Bundesministerium f\"ur Bildung und 
Forschung (BMBF) project no.\ 06 ER 808,
by Gesellschaft f\"ur Schwerionenforschung (GSI),
and by Graduiertenkolleg Schwerionenphysik.
The Joint Institute for Heavy Ion Research has as member institutions 
the University of Tennessee, Vanderbilt University, and the Oak Ridge 
National Laboratory; it is supported by the members and by the Department 
of Energy through Contract No.\ DE--FG05--87ER40361 with the University of 
Tennessee. 
%
%========================================================================
%
\begin{appendix}
\section{Details of the Mean--Field Models}
\label{Sect:Parameters}
\subsection{Skyrme Energy Functional}
\label{Subsect:SHF}
The Skyrme forces are constructed to be effective forces for
nuclear mean--field calculations. In this paper, we use the Skyrme 
energy functional in the following form
\begin{displaymath}
{\cal E}= {\cal E}_{\rm kin} [\tau ]
      + {\cal E}_{\rm Sk}  [\rho,\tau, {\bf J}]
      + {\cal E}_{\rm C}   [\rho_p]
      - {\cal E}_{\rm c.m.}
\end{displaymath}
with
\begin{eqnarray*}
{\cal E}_{\rm Sk} 
& = & \int \! {\rm d}^3r 
      \bigg(
            \frac{b_0}{2}\rho^2
          - \frac{b'_0}{2} \sum_q \rho_q^2
          + \frac{b_3}{3} \rho^{\alpha +2}
          - \frac{b'_3}{3} \rho^\alpha \sum_q \rho_q^2
      \nonumber \\
&   & \quad 
          + b_1  \rho \tau
          - b'_1 \sum_q \rho_q \tau_q
          - \frac{b_2}{2}  \rho \Delta \rho 
          + \frac{b'_2}{2} \sum_q \rho_q \Delta \rho_q
      \nonumber \\
&   & \quad
          - b_4  \rho \nabla \cdot {\bf {J}}
          - b'_4 \sum_q \rho_q \nabla \cdot {\bf J}_q
      \bigg)
\end{eqnarray*}
and \mbox{$q \in \{ {\rm p,n } \}$}. $\rho_q$, $\tau_q$, and ${\bf J}_q$
denote the local density, kinetic density, and spin--orbit current,
that are given by
\begin{eqnarray*}
\rho_q
& = & \sum_{k \in \Omega_q} v_k^2  | \psi_k |^2, \quad
\tau_q=   \sum_{k \in \Omega_q} v_k^2  | \nabla \psi_k |^2, \nonumber \\
{\bf J}_q
& = & - {\textstyle\frac{i}{2}} \sum_{k \in \Omega_q} v_k^2 
      \Big[ \psi_k^\dagger \, \nabla \times \hat\sigma \, \psi_k
            - ( \nabla \times \hat\sigma \, \psi_k )^\dagger \psi_k
      \Big].
\end{eqnarray*}
Densities without index denote total densities, e.g.\ 
\mbox{$\rho = \rho_{\rm p} + \rho_{\rm n}$}. The $\psi_k$
are the single--particle wavefunctions and $v_k^2$ the
occupation probabilities calculated taking the residual
pairing interaction into account, see Appendix~\ref{Subsect:PairFunc}.
${\cal E}_{\rm kin}$ is the kinetic energy 
\mbox{${\cal E}_{\rm kin} = [\hbar^2/(2m)] \int \! {\rm d}^3 r \, \tau$}, while
${\cal E}_{\rm C}$ is the Coulomb energy including the exchange term 
in Slater approximation. The center--of--mass correction reads 
\begin{equation}
\label{eq:Ecm}
{\cal  E}_{\rm c.m.}
= \frac{1}{2mA} 
   \bigl\langle\hat\textbf{P}\rule{0pt}{6.8pt}^2_{\rm c.m.}\bigr\rangle,
\end{equation}
where $\hat\textbf{P}_{\rm c.m.}$ is the total momentum operator in
the center--of--mass  
frame. The correction is calculated perturbatively by subtracting 
(\ref{eq:Ecm}) from the Skyrme functional after the convergence of the 
Hartree--Fock iteration.
The parameters $b_i$ and $b'_i$ used in the above definition
are chosen to give a compact formulation of the energy functional,
the corresponding mean--field Hamiltonian and residual interaction
\cite{Rei92b}. They are related to the more commonly used Skyrme force 
parameters $t_i$ and $x_i$ by
\begin{eqnarray*}
   b_0 & = &   t_0 \big( 1 + {\textstyle \frac{1}{2}} x_0 \big),\nonumber\\
   b'_0 & =&   t_0 \big( {\textstyle \frac{1}{2}} + x_0 \big),  
               \nonumber \\
   b_1 & = &  {\textstyle \frac{1}{4}} 
              \big[  t_1 \big( 1 + {\textstyle \frac{1}{2}} x_1 \big)
                    +t_2 \big( 1 + {\textstyle \frac{1}{2}} x_2 \big)
              \big], \nonumber \\
   b'_1& = &  {\textstyle \frac{1}{4}} 
              \big[  t_1 \big({\textstyle \frac{1}{2}} + x_1 \big)
                    -t_2 \big({\textstyle \frac{1}{2}} + x_2 \big)
              \big]  , \nonumber \\
   b_2 & = &  {\textstyle \frac{1}{8}} 
              \big[ 3t_1 \big( 1 + {\textstyle \frac{1}{2}} x_1 \big)
                    -t_2 \big( 1 + {\textstyle \frac{1}{2}} x_2 \big)
              \big]  , \nonumber \\
   b'_2& = &  {\textstyle \frac{1}{8}} 
              \big[ 3t_1 \big( {\textstyle \frac{1}{2}} + x_1 \big)
                    +t_2 \big( {\textstyle \frac{1}{2}} + x_2 \big)
              \big]  , \nonumber \\
   b_3 & = &  {\textstyle \frac{1}{4}} 
              t_3 \big( 1 + {\textstyle \frac{1}{2}} x_3 \big),\nonumber\\
   b'_3 & =&    {\textstyle \frac{1}{4}} 
              t_3 \big( {\textstyle \frac{1}{2}} + x_3 \big).
\end{eqnarray*}
The actual parameters for the parametrisation SkI4 used in this paper
are
\begin{displaymath}
  \begin{array}{rclrcl}
   t_0  & = & -1855.827 \,{\rm MeV}\,{\rm fm}^3, &
   x_0  & = & 0.405082,  \\
   t_1  & = & 473.829 \,{\rm MeV}\,{\rm fm}^5, &
   x_1  & = & -2.889148, \\ 
   t_2  & = & 1006.855 \,{\rm MeV}\,{\rm fm}^5, &
   x_2  & = & -1.325150, \\
   t_3  & = & 9703.607 \,{\rm MeV}\,{\rm fm}^{3+\alpha}, \  &
   x_3  & = & 1.145203, \\
   b_4  & = & 183.097  \, {\rm MeV} \, {\rm fm}^5 , &
   b'_4 & = & -180.351 \, {\rm MeV} \, {\rm fm}^5
  \end{array}
\end{displaymath}
with ${\alpha=0.25}$.
For the nucleon mass we use the value that gives
\mbox{$\hbar^2/(2m_{\rm p}) = \hbar^2/(2m_{\rm n})
= 20.7525 \, {\rm MeV} \, {\rm fm}^2$} for the constant entering 
${\cal E}_{\rm kin}$.
%
%----------------------------------------------------------------------------
%
\subsection{Relativistic Mean--Field Model}
\label{Subsect:RMF}
For the sake of a covariant notation, it is better to provide the
basic functional in the relativistic mean--field model as an
effective Lagrangian density, which for this study is defined as
\begin{displaymath}
{\cal L}_{\rm RMF} 
=  {\cal L}_{\rm N} + {\cal L}_{\rm M} + {\cal L}_{\rm NM} 
      + {\cal L}_{\rm em}, 
      % - {\cal E}_{\rm c.m.} 
\end{displaymath}
where
\begin{eqnarray*}
{\cal L}_{\rm M} 
& = & {\textstyle\frac{1}{2}}
      (\partial_\mu \Phi_\sigma \partial^\mu \Phi_\sigma 
      - {\cal U}_{\rm nonl}) 
      \\ 
&   & 
      - {\textstyle\frac{1}{2}}
        \Big[ {\textstyle\frac{1}{2}}
              (\partial_\mu\Phi^{\mbox{}}_{\omega,\nu}
               -\partial_\nu\Phi^{\mbox{}}_{\omega,\mu})
              \partial^\mu\Phi_\omega^{\nu}
              - m_\omega^2 \Phi_{\omega,\mu}^{\mbox{}} \Phi_\omega^\mu
        \Big]
       \nonumber \\ 
&   & 
      - {\textstyle\frac{1}{2}} 
        \Big[ {\textstyle\frac{1}{2}}
              (\partial_\mu\vec{\Phi}^{\mbox{}}_{\rho,\nu}
               -\partial_\nu\vec{\Phi}^{\mbox{}}_{\rho,\mu})\cdot
               \partial^\mu\vec{\Phi}_\rho^{\nu}
              - m_\omega^2 \vec{\Phi}_{\rho,\mu}^{\mbox{}}\cdot
             \vec{\Phi}_\rho^\mu
        \Big],\nonumber\\ 
{\cal L}_{\rm NM} 
& = & - g_\sigma \Phi_\sigma \rho^{\rm s}
      - g_\omega \Phi_{\omega,\mu} \rho^{\mu} 
      - g_\rho \vec{\Phi}_{\rho,\mu}\cdot\vec{\rho}^{\,\mu}, 
\\ 
{\cal U}_{\rm nonl}
& = & {\textstyle \frac{1}{2}} \Delta m^2 
      \biggl( \delta\Phi^2
              \log \biggl[\frac{\delta\Phi^2 + (\Phi_\sigma-\Phi_0)^2}
                               {\delta\Phi^2 + \Phi_0^2} \biggr]
     \nonumber \\
&   & 
\qquad\qquad \quad     + \frac{2 \Phi_0 \; \delta\Phi^2 \; \Phi_\sigma}
                     {\delta\Phi^2 + \Phi_0^2}
      \biggr)
      +{\textstyle\frac{1}{2}} m_\infty^2 \Phi_\sigma^2,
\\ 
{\cal L}_{\rm em} 
& = & - {\textstyle\frac{1}{2}} 
      (\partial_\mu A_\nu-\partial_\nu A_\mu) A^{\mu\nu} 
      - e A_\mu \rho_{\rm p}^\mu,
\end{eqnarray*}
and ${\cal L}_{\rm N}$ is the free Dirac Lagrangian for the nucleons
with nucleon mass \mbox{$m_{\rm N}=938.9\,{\rm MeV}$}, equally for protons and
neutrons.  The model includes couplings of the scalar--isoscalar
($\Phi_\sigma$), vector--isoscalar $(\Phi_{\omega,\mu}$),
vector--isovector ($\vec{\Phi}_{\rho,\mu}$), and electromagnetic ($A_\mu$)
field to the corresponding scalar--isoscalar ($\rho^{\rm s}$),
vector--isoscalar ($\rho^\mu$) and vector--isovector
($\vec{\rho}^{\,\mu}$) densities of the nucleons.  ${\cal U}_{\rm nonl}$
is the stabilized self--interaction of the scalar--isoscalar field, 
behaving like the standard {\em ansatz} for the nonlinearity at typical
nuclear scalar densities, but with an overall positive--definite
curvature to avoid instabilities at high scalar densities
\cite{Rei89,PL40}.  The actual parameters of the parametrisation
PL--40 are 
\begin{displaymath}
  \begin{array}{rclrcl}
  g_\omega   & = & 12.8861, &
  m_\omega   & = & 780.0 \, {\rm MeV}, \\
  g_\rho     & = & 4.81014, &
  m_\rho     & = & 763.0 \, {\rm MeV}, \\
  g_\sigma   & = & 10.0514, & \\
  m_\infty^2 & = & 4.0 \, {\rm fm}^{-2}, &
  \Delta m^2 & = & 3.70015 \, {\rm fm}^{-2}, \\
  \Phi_0     & = & -0.111914 \, {\rm fm}^{-1},\ &
  \delta\Phi & = & 0.269688  \, {\rm fm}^{-1}
  \end{array}
\end{displaymath}
(we follow the usual convention \mbox{$\hbar=c=1$} such that
\mbox{$197.3\,{\rm MeV}\equiv 1\,{\rm fm}^{-1}$}).
For the residual pairing interaction and the center--of--mass correction 
the same nonrelativistic approximations are used as in the SHF model.
%
%-----------------------------------------------------------------------
%
\subsection{Pairing Energy Functional}
\label{Subsect:PairFunc}
Pairing is treated in the BCS approximation using a delta pairing force
\cite{Krieger}, leading to the pairing energy functional
\begin{equation}
{\cal E}_{\rm pair}
= \frac{1}{4} \sum_{q \in \{ { \rm p, n } \} } V_q 
  \int \! {\rm d}^3 r \; \chi_q^2
\quad ,
\end{equation}
where \mbox{$\chi_q = - 2 \sum_{k \in \Omega_q > 0} f_k u_k v_k \; | \psi_k |^2$}
is the pairing density including state--dependent cutoff factors
$f_k$ to restrict the pairing interaction to the vicinity of the
Fermi surface \cite{pairStrength}. $v_k^2$ is the occupation probability  
of the corresponding single--particle state and $u_k^2 = 1 - v_k^2$. 
The strengths $V_{\rm p}$ for protons and $V_{\rm n}$ for neutrons depend 
on the actual mean--field parametrisation. They are optimized by fitting 
for each parametrisation separately the pairing gaps in isotopic and
isotonic chains of semi--magic nuclei throughout the chart of
nuclei. The actual values are
\begin{displaymath}
V_{\rm p} = -310 \, {\rm MeV} \, {\rm fm}^3, \quad
V_{\rm n} = -323 \, {\rm MeV} \, {\rm fm}^3
\end{displaymath}
in case of SkI4 and
\begin{displaymath}
V_{\rm p} = -348 \, {\rm MeV} \, {\rm fm}^3, \quad
V_{\rm n} = -346 \, {\rm MeV} \, {\rm fm}^3
\end{displaymath}
for PL--40.
The pairing--active space $\Omega_q$ is chosen to include one additional 
oscillator shell of states above the Fermi energy with a smooth Fermi 
cutoff weight, for details see \cite{pairStrength}.
\end{appendix}
%
%========================================================================
%

\end{document}